\pdfoutput=1

\documentclass[reprint,amsmath,amssymb,floatfix,longbibliography,aip,letterpaper,cha]{revtex4-1}

\usepackage{graphics,graphicx,float}

\makeatletter
\preprintsty@sw{}{%
 \usepackage{titlesec}
 \titleformat{\section}{\normalfont\footnotesize\sffamily\bfseries\uppercase}%
 	{\thesection}{1em}{}%
 \titleformat{\subsection}{\normalfont\small\sffamily\bfseries}%
 	{\thesubsection}{1em}{}%
 \titleformat{\subsubsection}{\normalfont\small\sffamily\slshape}{\thesubsubsection}{1em}{}%
 }%
\makeatother
  
\def\citeyear{\citep}
\def\autocite{\citep}

\usepackage[pdftex,
            pdfauthor={Steven A. Frank},
            pdftitle={Universal expressions of population change by the Price equation: natural selection, information, and maximum entropy production},
            pdfsubject={Universal expressions of population change by the Price equation: natural selection, information, and maximum entropy production}]{hyperref} 

\usepackage[dvipsnames]{xcolor}
\hypersetup{
    colorlinks = true,
    linkcolor = {black},
    urlcolor = {Mahogany},
}

\newcommand{\bdq}{\GD\mathbf{q}}
\newcommand{\bq}{\mathbf{q}}
\newcommand{\bdqs}{\bdq^{\mathbf{*}}}
\newcommand{\Dq}{\GD q}
\newcommand{\bda}{\GD\mathbf{a}}
\newcommand{\ba}{\mathbf{a}}
\newcommand{\bOne}{\mathbf{1}}
\newcommand{\bdw}{\GD\mathbf{w}}
\newcommand{\bw}{\mathbf{w}}
\newcommand{\obq}{\mathbf{\dot{q}}}
\newcommand{\oq}{{\dot{q}}}

\newcommand{\qhati}{{\hat{q}_i}}
\newcommand{\bqhat}{{\mathbf{\hat{q}}}}

\newcommand{\tq}{{\tilde{q}}}
\newcommand{\tw}{{\tilde{w}}}
\newcommand{\tm}{{\tilde{m}}}

\newcommand{\bdz}{\GD\mathbf{z}}
\newcommand{\bz}{\mathbf{z}}
\newcommand{\obz}{\mathbf{\dot{z}}}

\newcommand{\bdm}{\GD\mathbf{m}}
\newcommand{\bmm}{\mathbf{m}}

\newcommand{\zbar}{\bar{z}}
\newcommand{\mbar}{\bar{m}}
\newcommand{\abar}{\bar{a}}
\newcommand{\wbar}{\bar{w}}
\newcommand{\ozbar}{\dot{\bar{z}}}

\newcommand{\bF}{\mathbf{F}}
\newcommand{\bI}{\mathbf{I}}
\newcommand{\tF}{\tilde{\mathbf{F}}}
\newcommand{\tA}{\tilde{\mathbf{A}}}
\newcommand{\tI}{\tilde{\mathbf{I}}}

\newcommand{\Gfbh}{\boldsymbol{\hat{\Gf}}}
\newcommand{\Gfh}{\hat{\Gf}}
\newcommand{\Gfbar}{{\bar{{\Gf}}}}

\newcommand{\LL}{\mathcal{L}}
\newcommand{\EE}{\mathcal E}
\newcommand{\KL}[2]{\mathcal D\left(#1||#2\right)}
\newcommand{\D}{\mathcal D}
\newcommand{\J}{\mathcal J}
\newcommand{\F}{\mathcal F}
\newcommand*{\Gx}{\xi}

\newcommand*{\GF}{\Phi}
\newcommand{\tr}{T}

\newcommand*{\lr}[1]{\left(#1\right)}

\newcommand*{\Gb}{\beta}

\newcommand*{\GD}{\Delta}

\newcommand*{\Gk}{\kappa}
\newcommand*{\Gl}{\lambda}

\newcommand*{\Gm}{\mu}

\newcommand*{\Gs}{\sigma}

\newcommand*{\Gth}{\theta}
\newcommand*{\Gf}{\phi}

\newcommand*{\dd}{\textrm{d}}

\newcommand*{\Eq}[1]{eqn~\ref{eq:#1}}

\newcommand*{\ovr}[2]{{{#1}\over{#2}}}

\newcommand*{\prt}{\partial}
\newcommand*{\povr}[2]{\ovr{\prt #1}{\prt #2}}

\newcount\BoxNum \BoxNum 1\relax
\makeatletter
\newcommand*{\boxlabel}[1]{%
  \protected@write \@auxout {}{\string \newlabel {box:#1}{{\the\BoxNum}}{}}%
  \advance\BoxNum 1\relax}
\makeatother

\usepackage[numbib,notindex,nottoc,notlof,notlot]{tocbibind}
\settocbibname{References}

\newif\ifshowgit \showgittrue		% switches footer on/off
\newif\ifshowtime \showtimetrue		% show current time 
\ifshowgit
\usepackage[calc,timezonesep=, showseconds=false]{datetime2} 	
\DTMsavenow{now}
\DTMtozulu{now}{zdate}			% use \DTMuse{zdate} or \DTMuse{now} for Z zone or local zone
\usepackage[grumpy,local,mark,dirty=-*,raisemark=0.02\paperheight]{gitinfo2}

\fi

\usepackage{titlesec}
\renewcommand{\thesection}{\arabic{section}}
\renewcommand{\thesubsection}{\thesection.\arabic{subsection}}
\renewcommand{\thesubsubsection}{\thesubsection.\arabic{subsubsection}}

\begin{document}

\title{Universal expressions of population change by the Price equation: natural selection, information, and maximum entropy production\\ \phantom{x}}

%Alt title: Universal expressions of population change by the Price equation: natural selection, information, and a partial maximum entropy production principle

\author{Steven A.\ Frank}
%\email[email: ]{safrank@uci.edu}
%\homepage[homepage: ]{https://stevefrank.org}
\affiliation{Department of Ecology and Evolutionary Biology, University of California, Irvine, CA 92697--2525  USA}

\begin{abstract}

The Price equation shows the unity between the fundamental expressions of change in biology, in information and entropy descriptions of populations, and in aspects of thermodynamics. The Price equation partitions the change in the average value of a metric between two populations. A population may be composed of organisms or particles or any members of a set to which we can assign probabilities. A metric may be biological fitness or physical energy or the output of an arbitrarily complicated function that assigns quantitative values to members of the population. The first part of the Price equation describes how directly applied forces change the probabilities assigned to members of the population when holding constant the metrical values of the members---a fixed metrical frame of reference. The second part describes how the metrical values change, altering the metrical frame of reference. In canonical examples, the direct forces balance the changing metrical frame of reference, leaving the average or total metrical values unchanged. In biology, relative reproductive success (fitness) remains invariant as a simple consequence of the conservation of total probability. In physics, systems often conserve total energy. Nonconservative metrics can be described by starting with conserved metrics, and then studying how coordinate transformations between conserved and nonconserved metrics alter the geometry of the dynamics and the aggregate values of populations. From this abstract perspective, key results from different subjects appear more simply as universal geometric principles for the dynamics of populations subject to the constraints of particular conserved quantities\footnote{Published version: \href{http://dx.doi.org/10.1002/ece3.2922}{doi:\ 10.1002/ece3.2922} in \textit{Ecology \& Evolution}}\footnote{web: \href{https://stevefrank.org}{https://stevefrank.org}}.  

\bigskip


\end{abstract}

\maketitle

{\renewcommand{\tocname}{}\small\hbox{\null}\vskip-66pt\tableofcontents}\newpage

%%%%%%%%%%%%%%%%%%%%%%%%%%%%%%%%%%%%%%%%%%%%%%%%%%%%%%%%

\section{Introduction}

Changes in populations can often be described by changes in probability distributions. The dynamics of probability distributions therefore sets the basis for much of theoretical population biology. 

This article develops abstract principles for the dynamics of probability distributions. Those abstract principles deepen general understanding, leading to better connections of theoretical population biology to physics, statistics, and other population based disciplines. 

To understand the dynamics of probability distributions, one must consider the forces and constraints that influence the change in populations. Many methods can be used to study dynamics. Here, I apply the Price equation, a highly abstract description of change in populations. The abstractness of the Price equation facilitates discovery and understanding of connections between seemingly different disciplines. 

I use the Price equation to show the essentially identical basis for fundamental equations of natural selection, entropy and information. I emphasize the first steps in how one might go about building a common framework in which to understand the similarities and differences between various disciplines. From this abstract perspective, key results from different subjects appear more simply as universal geometric principles for the dynamics of populations subject to the constraints of particular conserved quantities.

\section{Overview}

This article provides the basis for unifying diverse subjects. Given the incompatible goals, methods, languages and cultures of the different disciplines, it is useful to begin with an extended overview. 

This overview serves only to orient in the direction of what follows, not as a complete summary unto itself. Readers who prefer to start with the details may wish to skip this section.

Sections 3--5 introduce the Price equation and prepare for application to different subjects. In the Price equation, a population consists of different types. Each type associates with a frequency or probability and with a property. I assume that the properties are quantitative values.  I use the words \textit{frequency} and \textit{probability} interchangeably. In other contexts, there may be good reasons to distinguish between these words.

The Price equation partitions the total change between two populations into a part caused by changes in frequencies and a part caused by changes in properties. That separation allows clear understanding of dynamics in terms of changes in probability distributions and changes in population quantities, such as biological fitness or physical energy or economic wealth.

Section 6 presents the canonical equation of conservation in populations, in which the change caused by frequency differences balances the change caused by property value differences. In biology, this equation represents the fact that the average of relative reproductive success (fitness) cannot change, because increases in relative fitness caused by natural selection must be exactly balanced by decreases in relative fitness caused by the changed state of the population. 

The conservation of relative fitness arises directly from the conservation of total probability. Alternative measures of property values can be understood as geometric coordinate transformations from the property of fitness (frequency change) to alternative measures that often lead to nonconservative changes in populations. For example, a logarithmic measure of fitness leads to classical measures of information.

Section 7 describes various identities and alternative partitions for the conservation of total probability. The different notational forms provide the basis for connecting seemingly different subjects to the common underlying geometric principles. 

Section 8 considers frequency changes in relation to an abstract notion of force. By expressing frequency changes in terms of force, the Price equation partitions the conservation of total probability into two balancing components of change. The first component arises from directly acting forces with respect to a fixed frame of reference for the quantitative properties. The second balancing component of change arises from the inertial forces that alter the frame of reference. 

The balance between the consequences of the direct and inertial forces provides an analogy to d'Alembert's principle of mechanics. That connection establishes a first step in relating different disciplines to the common underlying geometric foundation.

Sections 9--11 transform the quantitative property of frequency change into logarithmic coordinates. In the canonical Price equation's partition of conserved total probability into direct and inertial components, the property of each type is its frequency change or growth rate, an analogy with biological fitness. In particular, the relative growth, or fitness, of the $i$th type is $w_i=q_i'/q_i$, the ratio of the derived frequency, $q_i'$, relative to the initial frequency, $q_i$.

The change between the initial and derived frequency can be considered as a path divided into segments, in which the overall growth, or fitness, arises by multiplication of the fitnesses along each segment of the path. 

If we transform our focal property of fitness to logarithmic coordinates, then we can add component property values along the segments of a path, achieving an additive geometry of change that greatly enhances the power of analysis and interpretation. The classical notions of information and entropy follow immediately from use of the logarithmic coordinates in the canonical Price equation partition of conserved total probability.

Sections 12 and 13 continue to set the geometric foundations for analysis. When we divide a path of change into many small segments, then we can think of overall change as the combination of many small instantaneous changes in response to directly applied force at each point along the path. 

For small changes, the direct force at each point becomes approximately the same for the initial linear coordinates of change, $w_i$, and the logarithmic coordinates, $\log w_i$, apart from a constant shift that does not alter the dynamics. The convergence of linear and logarithmic coordinates with respect to small changes explains the common forms of many fundamental results in different fields of study.

Section 14 develops two complementary abstract notions of force. In the canonical expression of the Price equation for the conservation of total probability, the ``fitness'' term $w_i=q_i'/q_i$ simply describes the change in frequencies relative to the fixed frame of reference given by the initial frequencies. One may treat this description of change as an inductive expression of an underlying force. 

Alternatively, it often makes sense to consider the initial frequencies and forces as given, from which one deduces the change in frequency. This section expresses the given forces geometrically by the separation between the initial frequencies, $q_i$, and the given point, $\qhati$. By expressing force in this way, we have a common geometric basis for the inductive and deductive perspectives.

Section 15 develops the deductive perspective by deriving the changes in frequencies for given initial frequencies and given forces. The analysis applies the Lagrangian method, which maximizes the first component of the Price equation partition. That first component is an abstraction of the classical mechanics action term, as the virtual work of the direct forces with respect to a fixed frame of reference. The Lagrangian method generalizes the principle of least action.

The Lagrangian also includes various forces of constraint, such as the conservation of total probability, and any additional forces associated with other conserved quantities. The forces of constraint impose a limited set of potential paths that may be followed in the geometric space of frequency change. The actual path of change extremizes the action among those paths that are consistent with the forces of constraint.

Sections 16--18 present a partial maximum entropy production principle that follows from the dynamics of frequency change. To obtain this result, I partition the direct force into two components. The first component becomes an additional force of constraint that expresses the invariance imposed by the conservation of some system quantity, such as energy or biomass or the direct change in some value. The remaining component of the direct force is $-\log q_i$, which can be thought of as the entropy or information in the $i$th dimension. 

The entropy becomes the action term maximized by the path of change, leading to a path that maximizes the production of entropy. Because the maximization is taken with respect to the fixed frame of reference defined by the initial population, ignoring any inertial forces that alter the frame of reference, one can think of the entropy production as the result of a partial change holding constant the frame of reference---the partial maximum entropy production principle.

Sections 19 and 20 develop the notion of a conserved system quantity as a force of constraint. Jaynes  maximum entropy analysis of thermodynamics and probability patterns follows as a special case of the general geometric principles of change in populations developed in earlier sections. From Jaynes' work and the later extensions of his theory to simple invariance principles, we have a unified framework in which to understand the relations between commonly observed probability distributions. 

Section 21 discusses alternative ways in which to interpret maximum entropy paths. I argue that the most basic principles derive from the underlying geometry. Notions of entropy and information are simply interpretations of that geometry applied to particular disciplines of study.

Section 22 relates the path of change for populations to the Fisher information metric. That metric arises frequently in particular disciplines, including the fundamental approaches of information geometry. 

Sections 23 and 24 briefly review key results. The Appendix provides brief histories of key topics and background references.

\section{Separation of frequency and property}

The Price equation provides an abstract way in which to analyze changes in populations. The equation separates the frequency of entities from the property of those entities\autocite{price72extension,frank12naturalb}. 

Suppose, for example, that for entities with label $i$, we express frequency as $q_i$ and the average of the associated property value as $z_i$. The $z_i$ values can be height, or energy level, or any quantity. 

If entities with label $i$ always have an average value, $z_i$, then frequency change completely describes population change. If the change in frequency between two populations is $\GD q_i=q_i' - q_i$, then the change in the average value of $z$ is
\begin{equation*}
  \GD\zbar = \sum q_i'z_i - q_iz_i =\sum\GD q_i z_i = \bdq\cdot\bz,
\end{equation*}
in which the dot product, $\bdq\cdot\bz$, is understood in the usual way as the sum of the element-wise product of two vectors. 

Alternatively, one may separate frequency from property. Thus, we have differences in frequency, $\GD q_i = q_i' - q_i$, and differences in property values, $\GD z_i = z_i' - z_i$. 

For example, a transportation planner might study the overall assessment of changing modes of transport in a population. The index $i$ could label different transportation modes, such as automobile, train, and so on. The frequency $q_i$ is the fraction of individuals who travel by a particular mode. The quantity $z_i$ may be the relative assessment for the value associated with a transportation mode. 

The separation of frequency and property allows a more general description of change. Changes in the total assessment of transportation can arise from changes in the frequencies of usage, $\GD q_i = q_i' - q_i$, and from changes in the assessment of value for each mode, $\GD z_i = z_i' - z_i$. 

\section{Set mapping of labels between populations}

Our goal is to describe the change between two populations. We may arbitrarily label one population as the ancestor and the second population as the descendant. The general formulation concerns only the differences between populations, independently of any particular underlying scale of separation, such as space or time or updating in light of new evidence. In this section, I consider the example of separation between populations by time.

The term $\GD q_i = q_i' - q_i$ is the change in the descendant frequency, $q_i'$, compared with the ancestral frequency, $q_i$.  For the transportation example, one would typically read this as the frequency of people traveling by train or other mode, $i$, at two different times. If the frequency of people traveling by train is increasing, then $\GD q_i$ is positive. That interpretation makes a lot of sense and is nearly universal. 

The Price equation allows a more abstract notion of the mapping between sets. Let $q_i'$ be the frequency of entities in the second population that derive from type $i$ in the first population. Thus, for travel mode by train, $q_i'$ would be the frequency of individuals in the descendant population who derived from, or map to, train travelers in the ancestral population. 

Consider two interpretations. First, $q_i'$ and $q_i$ could have their traditional meaning of the frequencies of train travelers at each point in time. For example, change may occur by social contagion, in which people become train travelers only by learning about trains from someone who already travels by train; an individual train traveler maps to self as a descendant train traveler. In this case, each descendant train traveler maps to a train traveler in the ancestral population. Positive $\GD q_i$ reflects growth of the $i$th class by successful recruitment.

In a second interpretation, we could map descendant individuals to their mothers. Then $\GD q_i$ has to do with the number of babies produced by each mother. In this case, a descendant's label $i$ is defined only by ancestral type. Descendants do not have their own types, only their mapping to an ancestral $i$. 

We handle the fact that descendants may use travel modes that differ from their mother by adjusting the change in property value, $\GD z_i = z_i' - z_i$. For mothers who travel by train, with property value $z_i$, their descendants have some average property value, $z_i'$, that accounts for both changes in travel mode by descendants and changes in property value associated with each travel mode.

In the general, abstract interpretation, the label $i$ applies only to the initial, or ancestral set. All entities from the second, or descendant, population map to ancestors, and thus derive their labels from their ancestors. We can use partial assignments, so that a descendant is made up of various fractions of ancestors, each descendant part accounted for separately by its assignment to an ancestral label, $i$. 

At first glance, this set mapping abstraction may seem rather complicated and obscure. However, its great power arises from the fact that nearly all studies of changes in populations can be described by specific mapping assumptions and associated interpretations. Thus, anything that we can prove about the general abstract setup applies to the very many apparently different special cases that arise in different applications.

\section{The Price equation}

The Price equation \autocite{price72extension,frank12naturalb} describes the change between two populations in the aggregate value of some property (this section is modified from ref.~\onlinecite{frank15dalemberts}). Each component of the population has a frequency weighting, $q$, and a property value, $z$. Begin with a discrete analog of the chain rule for differentiation of a product
\begin{align*}
  \GD(qz) &= (q+\GD q)(z+\GD z)-qz\\
          &= (\GD q)z+(q+\GD q)\GD z\\
          &= (\GD q)z+q'\GD z,
\end{align*}
in which $q'=q+\GD q$ and $z'=z+\GD z$. The same chain rule can be applied to vectors. Using dot product notation, we obtain an abstract form of the Price equation \autocite{frank12naturalb,frank12naturalc,frank13natural}
\begin{equation}\label{eq:priceGD}
  \GD(\bq\cdot\bz) = \bdq\cdot\bz + \bq'\cdot\bdz,
\end{equation}
in which a dot product is understood in the usual way as $\bq\cdot\bz=\sum q_iz_i$.

This equation can be interpreted in various ways, as discussed in prior sections. In general analysis, I adopt the most abstract interpretation with regard to set mapping between two populations. Roughly speaking, we can take $q_i$ to be the frequency associated with a subset, $i$, of the initial population, such that the total frequency is $\sum q_i=1$. Thus, $\zbar=\sum q_iz_i$ is the average of $z$.

Here, $z_i$ is an arbitrary function that maps $i$ to some property value, and $z_i$ is interpreted as the average of $z$ in each dimension or subset, $i$. Because $z$ can be any quantity, calculated in any way, this equation gives the most general expression for $\GD\zbar$, the change in the average of $z$. One can think of $\zbar=\sum q_i z_i$ as a functional of the arbitrary function, $z$, that maps $i\mapsto z_i$.

For a second population, with frequencies $q_i'$ and values $z_i'$, we have $\sum q_i'=1$, in which the primes denote the abstract mapping described in the prior section. Our only restriction is that we can map the index $i$ between the two populations. We may define the average value in the second population as $\zbar'=\sum q_i'z_i'$. Thus, 
\begin{equation*}
  \GD\zbar = \zbar'-\zbar = \GD(\bq\cdot\bz), 
\end{equation*}
so that we may write the Price equation in \Eq{priceGD} as
\begin{equation}\label{eq:priceDZ}
  \GD\zbar = \bdq\cdot\bz + \bq'\cdot\bdz,
\end{equation}
an explicit expression for the change in average values. Because $z$ can be defined in any way, this expression describes the change in any quantitative property of populations. 

\section{Biological fitness and the conservation of total probability}

We may define an abstract analog of biological fitness. For a type or subset with label $i$, comprising frequency $q_i$ in the ancestral population, the fraction of the descendant population derived from $i$ is $q_i'$. Thus, the relative success of type $i$ in contributing to the descendant population may be written as its relative fitness
\begin{equation}\label{eq:fitness}
  w_i=\frac{q_i'}{q_i}.
\end{equation}
Average relative fitness is always one
\begin{equation*}
  \wbar=\sum_i q_i w_i = \sum_i q_i\left(\frac{q_i'}{q_i}\right)=\sum_i q_i' = 1,
\end{equation*}
because the total frequency or probability is always a conserved value of one. In some articles, $w_i$ is taken as an absolute measure of the number of descendants assigned to type $i$, and $\wbar$ is the average number of descendants, which may differ from one. In that case, $w_i/\wbar$ is relative fitness. Here, I am using $w_i$ as the measure of relative fitness, with $\wbar$ always equal to one. The following analysis does not differ under the alternative definitions, but it is important to keep in mind the distinct definitions that may be used.

If we use relative fitness for the abstract property in the Price equation of \Eq{priceDZ}, with $z\mapsto w$, we obtain
\begin{equation}\label{eq:consW}
  \GD\wbar = \bdq\cdot\bw + \bq'\cdot\bdw = 0.
\end{equation}

It is often useful to express fitnesses as deviations from their average value, which we obtain by subtracting one from relative fitness
\begin{equation}\label{eq:aveEx}
  a_i = w_i - 1 = \frac{q_i'}{q_i} - 1 = \frac{\Dq_i}{q_i},
\end{equation}
which is known as Fisher's average excess in fitness\autocite{fisher58the-genetical}. The average value $a$ is always zero, thus we can write \Eq{consW} as
\begin{equation}\label{eq:consA}
  \GD\abar = \bdq\cdot\ba + \bq'\cdot\bda = 0.
\end{equation}

\section{Identities for the conservation of probability}

We may express the conservation of total probability in a variety of equivalent forms. This section shows some of the variants. The purpose of these variants is to set up the discussion in the next section, in which we interpret the Price equation partition in \Eq{consA} as a partition of total change into two parts. The first part is the change ascribed to direct forces, $\bF$. The second part is the change ascribed to the altered context of the population, which may be thought of as a change in the frame of reference caused by inertial forces, $\bI$. 

I will discuss the interpretation of direct and inertial forces in the next section. Here, we must first consider various notational manipulations, which by themselves do not have much obvious meaning. The goal will ultimately be to discuss general aspects of change in populations subject to the constraint set by the conservation of total probability, which allows us to write the Price equation partition in \Eq{consA} as
\begin{equation}\label{eq:dalemb1}
  \GD\abar = \left(\bF + \bI\right)\bdq = 0.
\end{equation}
We will need a toolkit of notational variants to establish this form and to show the connections between seemingly different subjects. It is a bit tedious to set up the various notational identities, but it is important to do so to develop alternative interpretations and to avoid confusion. On first reading, one may wish skim quickly through this section and then refer back to the notations as needed. 

To start, note that $\bq'=\bq+\bdq$ and $\bda=\ba'-\ba$, thus we can write the second term of \Eq{consA} as
\begin{equation}\label{eq:qpda}
  \bq'\cdot\bda = \bq'\cdot\ba' - \bq\cdot\ba - \bdq\cdot\ba = -\bdq\cdot\ba,
\end{equation}
because $\bq'\cdot\ba'=\bq\cdot\ba$ are the average values of $a$, which are always zero. Thus, we end up with the seemingly trivial partition
\begin{equation}\label{eq:consA2}
  \GD\abar = \bdq\cdot\ba - \bdq\cdot\ba = 0,
\end{equation}
which we will nonetheless find quite useful, because the partition provides some hints about the balance of direct and inertial forces in a conservative system. Before turning to that balance of forces in the next section, it is useful to consider some additional identities.

Each term in \Eq{consA2} expresses the variance in fitness and, equivalently, a measure of the squared Euclidean distance through which the population moves
\begin{equation*}
  \bdq\cdot\ba = \bq\cdot\ba^2=V_w,
\end{equation*}
in which $\ba^2$ is the vector of the squared terms, $a_i^2$, and thus $\bq\cdot\ba^2$ is the second moment of $a$. Here, $V_w$ is the variance in relative fitness, because $a_i=w_i-1$ is relative fitness shifted so that the mean value of $a$ is zero. Thus, the second moment of $a$ is the variance.

The term $\bq\cdot\ba^2$ can be thought of as a squared distance starting from an initial point at zero and moving through the distance given by the sum of the squared deviations in each dimension, $a_i^2$, each dimension weighted by its frequency, $q_i$. Thus, the distance that the population moves in frequency space, caused by the changes in frequency given by variable fitnesses, is equivalent to the variance in fitness. Put another way, the reason that the variance in fitness always arises as the key metric in population change is that the variance describes the distance that the population moves.

We can also write
\begin{equation*}
  \bdq\cdot\ba = \sum_i\frac{\left(\GD q_i\right)^2}{q_i} 
  	= \sum_i q_i\left(\frac{\GD q_i}{q_i}\right)^2,
\end{equation*}
which are forms that arise in information theory interpretations of frequency changes, and also clarify the geometric squared distance interpretation of frequency changes\autocite{amari00methods}. We can write this equation in a nonstandard vector notation, which will be convenient to use in this article, as
\begin{equation}\label{eq:dqa}
  \bdq\cdot\ba = \sum_i\frac{\left(\GD q_i\right)^2}{q_i} = \left(\frac{\bdq}{\bq}\right)\bdq,
\end{equation}
in which a ratio of vectors implies element-wise division, and vectors distribute through parentheses as dot products. 

We can also rewrite the second term of \Eq{consA} by rearranging \Eq{qpda} as
\begin{equation}\label{eq:qpda2}
   \bq'\cdot\bda = \left(\frac{\bdqs}{\bdq} - \ba\right)\bdq,
\end{equation}
in which 
\begin{equation*}
  \bdqs = \bq''- 2\bq'+\bq = \bdq'-\bdq,
\end{equation*}
which measures the nonlinearity, or bending, in the changes of $\bq$ in subsequent steps, which is roughly like an acceleration. 

Note that \Eq{qpda2} has $\GD q_i$ terms in the denominator, which may appear to be problematic when such terms include zero values. However, each term is always part of a dot product, yielding values of $\GD q_i^*$ for each term, thus we can always interpret such terms directly by their actual value. The reason for splitting the terms in the manner of \Eq{qpda2} follows at the end of this section.

Note also that
\begin{equation}\label{eq:middleterm}
  \frac{\bdqs}{\bdq}\cdot\bdq = \sum_i \GD q_i' - \GD q_i = 0
\end{equation}
by the conservation of total probability. However, in each individual dimension, $i$, the value of $\GD q_i^*=\GD q_i' - \GD q_i$ is not necessarily zero. Although the total value is constrained to be zero, it is often useful to retain this term to emphasize the fact that the values in each dimension can vary.

We can combine the various pieces to express the Price equation partition for the change in relative fitness in \Eq{consA} as
\begin{equation}\label{eq:dalembP1}
  \GD\abar = \left(\ba + \frac{\bdqs}{\bdq} - \ba\right)\bdq = 0,
\end{equation}
or, using $\ba=\bdq/\bq$, as
\begin{equation}\label{eq:dalembP2}
  \GD\abar = \left(\frac{\bdq}{\bq} + \frac{\bdqs}{\bdq} - \frac{\bdq}{\bq}\right)\bdq = 0.
\end{equation}
The second form emphasizes that this expression is given purely as the nondimensional description of changes in frequency or probability. Later, it will be useful to drop the middle term by using the identity in \Eq{middleterm}, leading to the form in \Eq{consA2} expressed as
\begin{equation}\label{eq:fisherdiscrete}
  \left(\frac{\bdq}{\bq} - \frac{\bdq}{\bq}\right)\bdq 
  	= \sum_i\frac{\lr{\GD q_i}^2}{q_i} - \frac{\lr{\GD q_i}^2}{q_i} = 0.
\end{equation}

\section{Balance of direct and inertial forces}

The previous sections described the conservation of total probability, which imposes strong constraints on the geometry of change in populations. In particular, the dynamics of probability distributions must move along the constraint that the total probability remains unchanged. Within that constraint, the probability distributions that characterize populations may change in response to directly applied forces, such as biological fitness or physical forces or informational processes. 

This section analyzes the changes in probability distributions in response to direct forces and subject to the constraint of conserved total probability. The previous section established the key equations. On the abstract side, \Eq{dalemb1} presented the partition between the forces that directly change frequencies, $\bF$, and the forces that change the inertial frame of reference for the population, $\bI$, as
\begin{equation*}
  \GD\abar = \left(\bF + \bI\right)\bdq = 0,
\end{equation*}
which expresses a nondimensional analogy of d'Alembert's principle with respect to the balance between the direct and inertial components\autocite{lanczos86the-variational}. D'Alembert's principle describes classical physical laws of motion in systems that conserve total energy, for example, motion that does not lose energy by friction and dissipation of heat. I previously discussed d'Alembert's principle in the context of frequency changes in populations\autocite{frank15dalemberts}. Here, I repeat a few key points from my previous article. 

The term $\bF$ is the vector of direct forces acting on the system, and the term $\bI$ is the vector of inertial forces that balance the direct forces to achieve no net change. d'Alembert's principle can be thought of as a generalization of Newton's second law of motion \autocite{lanczos86the-variational}, in which $\tF=\Gm\tA$ is read as the total force, $\tF$, equals mass, $\Gm$, times total acceleration, $\tA$. Total force and total acceleration must include forces of constraint, which in our case means that $\sum\GD q_i=0$. If we write total inertial force as $\tI =-\Gm\tA$, then Newton's law is $\tF+\tI=0$.

In d'Alembert's formulation, the direct and inertial forces typically do not sum to zero, $\bF+\bI=0$, because those terms do not include the constraining forces that act on $\bdq$. Instead, in d'Alembert's expression $\left(\bF + \bI\right)\bdq = 0$, the term $\bdq\cdot\bF$ combines the direct and constraining forces, and the term $\bdq\cdot\bI$ combines all inertial forces, including any forces of constraint. Newton's law is a special case of the more general principle of d'Alembert\autocite{lanczos86the-variational}.

Here is a simple intuitive description of d'Alembert's principle \autocite{wikipedia15fictitious}. You are sitting in a car at rest, and the car suddenly accelerates. You feel thrown back into the seat. But, even as the car gains speed, you effectively do not move in relation to the frame of reference of the car: your velocity relative to the car remains zero. That net zero velocity can be thought of as the balance between the direct force of the seat pushing on you and the inertial force sending you back as the car accelerates forward.

As long as your frame of reference moves with you, then your net motion in your frame of reference is zero. Put another way, there is a changing frame of reference that zeroes net change by balancing the work of direct forces against the work of inertial forces. Although the system is a dynamic expression of changing components, it also has an overall static, equilibrium quality that aids analysis. As Lanczos \autocite{lanczos86the-variational} emphasizes, d'Alembert's principle ``focuses attention on the forces, not on the moving body \dots''

In terms of explicit notation for changes in frequencies, the previous section developed a Price equation expression for the partition of direct and inertial forces in \Eq{dalembP2} as
\begin{equation}\label{eq:dalembP3}
  \GD\abar = \left(\frac{\bdq}{\bq} + \frac{\bdqs}{\bdq} - \frac{\bdq}{\bq}\right)\bdq = 0,
\end{equation}
with analogy to d'Alembert's form by expressing direct and inertial forces as
\begin{equation*}
  \bF = \frac{\bdq}{\bq} \mskip35mu \mathrm{and} 
  	\mskip35mu \bI = \frac{\bdqs}{\bdq} - \frac{\bdq}{\bq}.
\end{equation*}
For frequency changes, one can think of a coordinate system that locates a population as a point defined by the population's frequency or probability distribution. The direct work done to move the population in that coordinate system is $\bdq\cdot\bF$, the sum of the force multiplied by the displacement in each dimension, calculated when holding constant the frame of reference defined by the coordinate system. That direct work is balanced by the inertial work done to accelerate the reference frame coordinate system by a total amount $\bdq\cdot\bI$, which relocates the altered population and its associated forces so that it appears in the new frame of reference to have a net total displacement multiplied by force of zero.

I use the word ``force'' here in an abstract, nondimensional manner, rather than in the specifically defined manner of classical physics. Such words can be a barrier to interdisciplinary insight and understanding. Readers highly trained in particular disciplines, such as physics, sometimes believe that a word such as ``force'' has a single correct meaning and associated units of expression. Any variant use of the word is thought to be misleading or mistaken. I take the opposite view. The underlying nondimensional geometry expresses the purest abstract notion of such concepts. 

In each separate discipline, the particular dynamics and related equations have terms that take on specific interpretations, units, and meaning. Those specific aspects arise from the application of the same underlying universal geometry to particular problems, which usually means the same underlying conserved quantities and associated symmetries. The same geometry and abstract concepts will take on different units and interpretations in different disciplines.

\section{Average force along a path}

In the Price equation description of change, we have only the differences between two populations. The two populations describe the initial and final probability distributions, $\bq$ and $\bq'$. Each distribution can be thought of as a single point in a space of probability distributions. The separation between the two points is a nondimensional change that can be small or large. There is no underlying parameter, such as time or spatial distance, that defines the scale of separation and the path of change that connects the points.

Most applications analyze changes along a path with respect to an underlying parametric scale. To relate the Price equation to other theoretical frameworks, it is useful to add an abstract notion of change along a parametric path that connects the initial and final probability distributions. 

Let $\Gth$ be a parameter that describes change along a path that connects $\bq$ to $\bq'$
\begin{equation*}
  q_i'(\Gth) = q_i(\Gth_0)e^{r_i\GD\Gth},
\end{equation*}
in which $\GD\Gth=\Gth-\Gth_0$. We can set $\Gth_0=0$ and thus write $\Gth\equiv\GD\Gth$. For notational convenience, let the dependence of $q(\Gth)$ on the parameter $\Gth$ be implicit, so that we can write the same expression more simply as
\begin{equation}\label{eq:qdiffeq}
  q_i'=q_ie^{r_i\Gth}.
\end{equation}
We can think of $r_i$ as the average force acting along the path that moves the system from $q_i$ to $q_i'$ with respect to total path length, $\Gth=\GD s^2$, in the parametric length scale, $s$. Thus, $r_i\Gth$ is the total force in the $i$th dimension along the path of change. For our purposes, we can treat $s$ as a nondimensional scale, and think of $r_i$ as having nondimensional units of $1/s^2$, interpreted as a nondimensional force or acceleration. In biology, the force $r_i$ is interpreted as the Malthusian expression of biological fitness in analyses of natural selection, connecting the abstract analysis here to models of biological evolution \autocite{frank15dalemberts}.

Note that
\begin{equation*}
  r_i = \frac{1}{\GD\Gth}\log\frac{q_i'}{q_i}=\frac{\GD\log q_i}{\GD\Gth}.
\end{equation*}
so that we may think of $r_i$ as the average change in logarithmic coordinates of probability with respect changes in the parametric length scale $\GD\Gth=\GD s^2$. 

We can express the total nondimensional force in these logarithmic coordinates acting along the path of change from $\bq$ to $\bq'$ as
\begin{equation*}
  m_i = r_i\Gth = \log\frac{q_i'}{q_i}=\log w_i=\GD\log q_i.
\end{equation*}
Because $m=\log w$, we can think of $m$ as log fitness. Using $m$ to express fitness, or force, the expression for change along a path in \Eq{qdiffeq} becomes
\begin{equation*}
  q_i'=q_ie^{m_i}.
\end{equation*}

\section{Comparing linear and logarithmic coordinates}

In linear coordinates, for each implicit $i$, we combine forces multiplicatively
\begin{equation*}
  w=\frac{q'}{q}= \frac{q'}{\tq}\frac{\tq}{q}=\tw'\times \tw,
\end{equation*}
in which $\tq$ separates $\lr{q,q'}$ into the segments $\lr{q,\tq} + \lr{\tq,q'}$, with $\tq$ between $q$ and $q'$. 

In logarithmic coordinates, we combine forces additively
\begin{equation*}
     m=\log\frac{q'}{q}= \log\frac{q'}{\tq}\frac{\tq}{q}
     	=\log\frac{q'}{\tq}+\log\frac{\tq}{q}=\tm' + \tm.
\end{equation*}
The two coordinate systems describe the same total fitness, or force, as
\begin{equation}\label{eq:wmult}
	w=\frac{q'}{q} = e^{\tm'+\tm} = \tw' \times \tw.
\end{equation}
We can decompose any fitness value and its associated vector, $\lr{q,q'}$, into a large number of small pieces. In principle, we could analyze large changes in frequency, $\GD q = q'-q$, by combining the changes along each small segment in a decomposition of total change. 

\section{Log coordinates, entropy and information}

The average value of log fitness is
\begin{equation*}
  \mbar = \bq\cdot\bmm = \bq\cdot\log\frac{\bq'}{\bq}
  	= \bq\cdot\GD\log\bq=-\KL{\bq}{\bq'},
\end{equation*}
in which
\begin{equation*}
  \KL{\bq}{\bq'} = \sum_i q_i\log\frac{q_i}{q_i'}
\end{equation*}
is the Kullback-Leibler divergence\autocite{kullback59information,cover91elements}. This divergence measures relative entropy by extending the classical measure of entropy, $-\bq\cdot\log\bq$, for a probability vector $\bq$, to a measure of the entropic divergence of $\bq$ relative to a given probability vector, $\bq'$. 

One can think of classical entropy for a probability vector, $\bq$, as a special case of the more general relative entropy by comparing $\bq$ to a uniform distribution described by a constant probability vector in which $q_i'=1/N$ for all $i$. The Kullback-Leibler divergence is also a primary measure of information in statistics and information theory. 

The properties of entropy and information derive from the fundamental geometric properties of logarithmic coordinates, such as the additivity described in the previous section.

From the equality above, $\mbar=-\KL{\bq}{\bq'}$, we can write the change in mean log fitness as
\begin{equation*}
  \GD\mbar = \mbar'-\mbar = \KL{\bq}{\bq'} - \KL{\bq'}{\bq''},
\end{equation*}
which measures the bending, or curvature, of the divergence between the populations in the sequence $\bq\rightarrow\bq'\rightarrow\bq''$. When the divergence between successive steps remains constant, then mean log fitness is invariant. 

We can use the Price equation in \Eq{priceDZ} to partition the total change in log fitness into direct and inertial components
\begin{equation}\label{eq:priceDM}
  \GD\mbar = \bdq\cdot\bmm + \bq'\cdot\bdm.
\end{equation}
The direct component is
\begin{equation*}
  \bdq\cdot\bmm = \bdq\cdot\GD\log\bq = \KL{\bq}{\bq'} + \KL{\bq'}{\bq},
\end{equation*}
in which
\begin{equation}\label{eq:jeffreys}
  \J(\bq,\bq') = \KL{\bq}{\bq'} + \KL{\bq'}{\bq}
\end{equation}
is the Jeffreys divergence. In earlier work, I showed that the Jeffreys divergence is the proper expression for the direct component of change caused by natural selection or, more generally, the component associated with direct forces when evaluated with respect to the fixed frame of reference given by the initial probability vector\autocite{frank12naturalc}. 

For small changes, $\J$ and $\D$ converge to the Fisher information metric (see below). Thus, analyses of small changes often invoke $\J$, $\D$ or Fisher information without distinguishing between the measures. For small changes, the Fisher information metric is often preferable, because it has many useful geometric properties\autocite{amari00methods} and is more widely known than $\J$. However, it is useful to keep in mind that, in general, $\J$ is the correct measure for the direct effect of natural selection, or for the direct component of change relative to a fixed frame of reference. 

The inertial component is
\begin{equation*}
  \bq'\cdot\bdm = \GD\mbar - \J(\bq,\bq') = - \KL{\bq'}{\bq} - \KL{\bq'}{\bq''}.
\end{equation*}

\section{Small changes: prelude}

In the remainder of this article, I focus only on the small changes that arise from forces acting at a given point. Small changes correspond to a single small segment in any larger path. I focus on small changes for two reasons. 

First, the conceptual relations between different disciplines can be seen mostly clearly in small changes around a focal point. 

Second, analysis of larger changes requires either an assumed constancy of a force field, or potential function, or an explicit notion of how forces change with both time and the changing context of the population. Those required assumptions reduce the generality of any particular formulation and obscure the common conceptual basis of different subjects.

In the future, it would be useful to extend analysis to cases in which there is no meaningful decomposition of a large change vector into small segments and to cases in which there exists a constant force field for which one could reconstruct the path of change over a sequence of small segments. Such extensions exist within individual disciplines, but it remains unclear how to connect the analyses from those different subjects to a common unifying framework.

\section{Small changes: analysis}

When changes $\GD q_i=q_i'-q_i$ are small, I use the notation $\GD q_i\rightarrow\dd q_i\equiv\oq_i$. For linear coordinates, we may write
\begin{equation*}
  w_i = \frac{q_i'}{q_i} = 1+\frac{\GD q_i}{q_i} \rightarrow 1+\frac{\oq_i}{q_i},
\end{equation*}
and for logarithmic coordinates when $\oq_i/q_i$ is small, we may write
\begin{equation*}
  m_i =\log{w_i} \rightarrow \frac{\oq_i}{q_i}.
\end{equation*}
Because the consequence of forces is shift invariant in expressions such as
\begin{equation*}
  \bdq\cdot\bw \rightarrow \obq\cdot\bmm,
\end{equation*}
the linear and logarithmic expressions of force, $\bw$ and $\bmm$, are equivalent for small changes. We may express this equivalence explicitly by noting that, in general, the direct component of change was given earlier as
\begin{equation*}
  \bdq\cdot\bw = \bdq\cdot\frac{\bdq}{\bq} = \sum_i\frac{\lr{\GD q_i}^2}{q_i},
\end{equation*}
which, when $\oq_i/q_i$ is small, we may write as
\begin{equation*}
  \obq\cdot\bmm = \obq\cdot\frac{\obq}{\bq} = \sum_i\frac{\oq_i^2}{q_i}.
\end{equation*}
This last expression is the Fisher information metric, which arises as the direct component of population change or natural selection\autocite{frank09natural}, the limiting expression of the Jeffreys divergence given earlier.

\section{Given forces}

I have defined $m_i=\log{q_i'/q_i}\rightarrow\oq_i/q_i$ as proportional to the force acting along the infinitesimal change $\oq_i=q_i'-q_i$. These expressions describe a consistency relation between force and frequency change. Often, we wish to consider how extrinsic or given forces cause change, rather than simply express consistency. 

Suppose, for example, that we have a given force vector acting at the point in frequency space, $\bq$. The given force is the nondimensional vector 
\begin{equation}\label{eq:phihatdef}
  \Gfbh = \log\frac{\bqhat}{\bq}.
\end{equation}
Given the location, $\bq$, and the force vector, $\Gfbh$, the vector $\bqhat$ provides an alternative way to express the intensity of the force vector as $\log\bqhat/\bq$. We can multiple $\bqhat$ by an arbitrary positive constant, because the net consequences of a force vector are shift invariant. Thus, we may implicitly consider $c\,\bqhat$ as the target and choose $\bqhat$ to sum to one, satisfying the conservation of total probability.

As with $\bmm$, we can write the total nondimensional force as a description of an exponential growth process
\begin{equation*}
  \qhati = q_ie^{\Gfh_i},
\end{equation*}
in which $\qhati$ is the endpoint of the exponential growth process that began at $q_i$. Thus, the location $\bq$ and the ``target'' location $\bqhat$ are sufficient to describe the given force vector. In the following, we will only be interested in small changes, $\obq$, that result from the instantaneous given forces with respect to a fixed frame of reference. One goal will be to find the changes, $\obq$, that arise from given forces and various constraints on change.

It is common in classical mechanics to define force, $\Gfh_i$, in relation to coordinates, $q_i$, by the negative gradient of a potential function $\GF$, which for our definition of $\Gfbh$ leads to
\begin{equation*}
  \Gfh_i = -\povr{\GF}{q_i}= \log\frac{\qhati}{q_i}.
\end{equation*}
We can use the potential function
\begin{equation}\label{eq:potential}
  \GF = \KL{\bq}{\bqhat}-\left(\sum q_i -1\right),
\end{equation}
in which the second term expresses the constraint on total probability, so that the resulting force includes the force of constraint. The average force, $\bar{\Gf}=\bq\cdot\Gfbh=-\KL{\bq}{\bqhat}$, is also a relative entropy expression. 

\section{Extreme action and frequency dynamics}

The given forces and the conservation of total probability do not by themselves tell us what frequency changes occur. In the study of frequency changes, the simplest variational approach\autocite{lanczos86the-variational} finds the extremum (maximum or minimum) of a Lagrangian subject to a constraint. In our case, we may write
\begin{equation}\label{eq:lagr1}
  \LL=\sum_i\oq_i \Gfh_i-\frac{1}{2\Gk}\left(\sum_i\frac{\oq_i^2}{q_i}-C^2\right)
    -\Gx\left(\sum_i\oq_i-0\right),
\end{equation}
in which we take as given the direct force in each dimension, $\Gfh_i$. 

We measure the total change caused by the direct forces as $\obq\cdot\bmm=\sum\oq_im_i=\sum\oq_i^2/q_i$. That expression comes from Price's separation of direct and inertial forces in \Eq{priceDM}.  In terms of classical mechanics\autocite{lanczos86the-variational}, the expression $\obq\cdot\bmm$ is the virtual work of the direct forces, in which work is distance times force (ignoring mass). 

Geometrically, we can think of the constraint in the second term as fixing the total path length moved in frequency space\autocite{amari00methods}, in which $\sum \oq_i^2/q_i=C^2$ measures distance by the Fisher information metric for infinitesimal displacements, $\obq$, or, biologically, $C^2$ is the variance in fitness. I assume that $C^2$ is chosen so that a solution exists that satisfies the constraints. The final term constrains total probability to remain constant.

The constraints of $\sum\oq_i^2/q_i=C^2$ and $\sum\oq_i=0$ do not by themselves determine which frequency changes actually occur. Many different frequency vectors, $\obq$, satisfy those two constraints.

Given these forces and constraints, what actual path do the dynamics follow? In other words, what is the realized vector $\obq$? We can think of the first term in the Lagrangian as the action, and extremize the action subject to the given constraints\autocite{lanczos86the-variational}. That action term is $\obq\cdot\Gfbh$, the product of the displacement times the given force, which is the virtual work. In this case, maximizing the virtual work in the Lagrangian finds the displacement $\obq$ aligned with the direct and constraining forces. 

To find the extreme action path, we evaluate $\prt\LL/\prt \oq_i=0$, which yields 
\begin{equation}\label{eq:qdotphi}
  \oq_i=\Gk q_i\Gfh_i^*,
\end{equation}
in which $\Gfh_i^*=\Gfh_i-\Gfbar$ is the excess force relative to the average, and $\Gx=\Gfbar=\sum q_i\Gfh_i$ follows from satisfying the conservation of total probability and the assumption that the virtual displacements are small. The constant of proportionality 
\begin{equation}\label{eq:kappa}
  \Gk=\frac{C}{\Gs_{\Gfh}}
\end{equation}
satisfies the constraint on total path length, in which $\Gs_{\Gfh}$ is the standard deviation of the direct forces.

Here, we have deduced a fundamental expression for frequency dynamics by the principle of extreme action. We can rewrite the expression for frequency dynamics as
\begin{equation}\label{eq:mphi}
  m_i=\frac{\oq_i}{q_i}=\Gk \Gfh_i^*,
\end{equation}
which shows that the forces, $m_i$, may be arrived at inductively by consistency with given changes, $\oq_i/q_i$. This expression also shows that the forces described by $\bmm$ are related by affine transformation to a vector of given forces, $\Gfbh$, from which one may deduce the actual frequency changes.

\section{Direct forces and constraining forces}

The distinction between direct and constraining forces is arbitrary. We may choose to describe a force by its constraint on allowable displacements, $\obq$, or by its inclusion in the direct forces, $\Gfbh\equiv\bF$. 

The Lagrangian in \Eq{lagr1} defines the action to be extremized as the work done along the path, which is the total displacement, $\obq$, times the direct component of force, $\Gfbh$. We can use $\Gfbh$ rather than $\Gfbh^{\raisebox{-7pt}{*}}=\Gfbh-\Gfbar$ for force, because we can ignore the constant, $\Gk$, and $\obq\cdot\Gfbar=0$. 

The constraining forces in the Lagrangian of \Eq{lagr1} are the fixed path length, $\sum\oq_i^2/q_i=C^2$, and the conservation of total probability, $\sum\oq_i=0$.  

We are free to relabel a component of the direct force as a constraining force\autocite{lanczos86the-variational}. In practice, deriving the altered Lagrangian provides an easy way to see how the changed labeling of direct and constraining forces enters into the analysis. 

Consider the direct forces as defined in \Eq{phihatdef} as
\begin{equation*}
  \Gfbh = \log\frac{\bqhat}{\bq}=\log\bqhat-\log\bq.
\end{equation*}
We can think of this expression as the sum of two component forces, $\log\bqhat$ and $-\log\bq$. The virtual work term of the direct forces becomes
\begin{equation}\label{eq:split}
  \obq\cdot\Gfbh=\obq\cdot\log\bqhat-\obq\cdot\log\bq.
\end{equation}
We may choose to relabel $\obq\cdot\log\bqhat$ as a force of constraint. The remaining term $-\obq\cdot\log\bq$ becomes the virtual work associated with the direct forces. The next section illustrates how this change in labeling can be useful.

\section{Conserved system quantities as the primary forces of constraint}

In relabeling $\log\bqhat$ as a constraining force, we may write
\begin{equation}\label{eq:logqhat}
  \log\bqhat=\log k - \Gl\bz,
\end{equation}
in which $\log k$ is understood to be a constant vector with elements $k$ when used in a vector context, $k$ is chosen so that $\sum\qhati=1$ obeys the conservation of total probability, the term $\Gl$ is a positive constant, and $z_i>0$ is chosen to make the equality hold. Thus, we can express the force associated with $\qhati$ by using $z_i$. The constraining force now becomes associated with the component 
\begin{equation}\label{eq:zconstrain}
  \obq\cdot\log\bqhat=-\Gl\left(\obq\cdot\bz\right).
\end{equation}

The advantage of using $\bz$ is that we may define the force of constraint directly in terms of any system quantity that we may associate with $\bz$. Each $z_i$ is, in this analysis, a given value associated with a subset $i$ of the population. We can use any quantity for $\bz$, including energy or momentum or monetary wealth or a quantitative biological trait. 

Often, underlying quantities of a system, $x_i$, become transformed by various processes before we evaluate the final quantity of the outcome, $z_i$. We may, in general, consider $z_i=\tr(x_i)$, in which $x_i$ is an intrinsic quantitative value associated with the subset $i$, and $\tr(x_i)$ is a transformation that defines a scaling relation between the intrinsic $x_i$ values and the constraining force, $z_i$. The analysis of pattern often reduces to understanding the processes that set the scaling relation\autocite{frank14how-to-read}, $\tr$. 

Because we can define $z_i=\tr(x_i)$ in any way, the quantity $\zbar=\bq\cdot\bz$ can represent almost any sort of functional on the system. This expression for $\zbar$ is also the average value of $\bz$. It is often useful to consider changes in $\zbar$, with infinitesimal change as
\begin{equation}\label{eq:zprice}
  \ozbar=\obq\cdot\bz + \bq\cdot\obz,
\end{equation}
which we obtain by a simple chain rule expansion of the differential, yielding an infinitesimal expression of the Price equation given in \Eq{priceDZ}.

If $\obq\cdot\bz$ is constrained, then that constraint defines the constraint on $\bqhat$ in \Eq{zconstrain}. For example, the total system quantity $\zbar$ may be conserved, which means that $\ozbar=0$. If the $\bz$ quantities do not themselves change, then $\bq\cdot\obz=0$, and consequently we have the constraint on the given forces $\obq\cdot\bz=0$. We may also consider other ways in which $\obq\cdot\bz$ is constrained, thereby defining the given forces $\bqhat$ that determine dynamics.

\section{Maximum entropy production principle}

With the split between direct and constraining forces in \Eq{split}, and the expression of the constraining forces in terms of $\bz$ in \Eq{zconstrain}, we can write a new Lagrangian that is equivalent to the Lagrangian in \Eq{lagr1}, using dot product notation
\makeatletter
\preprintsty@sw%
{%
\begin{equation}\label{eq:lagr2}
  \LL=-\obq\cdot\log\bq-\frac{1}{2\Gk}\left(\obq\cdot\bmm-C^2\right)
    -\Gx\left(\obq\cdot\mathbf{1}-0\right)-\Gl\left(\obq\cdot\bz-B\right).
\end{equation}
}%
{%
\begin{equation}\label{eq:lagr2}
\begin{split}
  \LL=-\obq\cdot\log\bq-\frac{1}{2\Gk}\left(\obq\cdot\bmm-C^2\right)
    &-\Gx\left(\obq\cdot\mathbf{1}-0\right)\\
    &-\Gl\left(\obq\cdot\bz-B\right).
\end{split}
\end{equation}
}%
\makeatother
The first term is the total action to be maximized, which is the virtual work of the direct forces, $\obq\cdot\bF=-\obq\cdot\log\bq$. The other terms describe the constraints on the path that $\obq$ may follow. I assume that $C^2$ and $B$ are chosen such that a solution exists.

The classical definition of entropy is $-\bq\cdot\log\bq$. Thus, the path $\obq$ that maximizes $\obq\cdot\bF=-\obq\cdot\log\bq$, subject to the constraints on $\obq$, is, in the limit of small changes, the path that maximizes the production of entropy subject to the constraints---the maximum entropy production principle (see Appendix for references). 

The idea is that the most likely path is the one that maximizes the production of entropy, which is equivalent to the maximization of the virtual work of the direct forces, $\obq\cdot\bF=-\obq\cdot\log\bq$, subject to the constraints on $\obq$. The constraints in $\obq$ include all forces that determine the location of $\log\bqhat=\log k-\Gl\bz$.

The maximum entropy production principle is always true, in the sense that one can always split the total direct forces, $\Gfbh$, into a constraining component, $\log\bqhat$, and a direct component, $-\log\bq$. The extent to which maximum entropy production is meaningful depends on two questions. First, how meaningful is it to treat $\log\bqhat=\log k-\Gl\bz$ as a constraint? Second, how meaningful is it to consider paths of change in the context of the Price equation separation of direct and inertial forces, a generalization of d'Alembert's principle? 

In order to answer those questions about maximum entropy production, the next section analyzes dynamics with respect to $\bz$ as a constraint. The following section discusses the Jaynesian theory of maximum entropy in relation to equilibrium thermodynamic expressions for common probability distributions. After those two sections, I return to the broader question of how to interpret the maximum entropy production principle in terms of the Price equation.

\section{Maximum entropy path subject to constraint}

To interpret the meaning of $\bz$ as a constraint, we return to the Lagrangian in \Eq{lagr2}.
That Lagrangian is equivalent to the form in \Eq{lagr1}, thus solving $\prt{\LL}/\prt\oq_i=0$ yields a solution equivalent to \Eq{qdotphi}, which we can expand to emphasize alternative interpretations
\begin{equation*}
  \oq_i=\Gk q_i\Gfh_i^*=\Gk q_i\left(\EE_i^*-\Gl z_i^*\right),
\end{equation*}
with deviations from average values $z_i^*=z_i-\zbar$ and
\begin{equation*}
  \EE_i^*=\EE_i-\EE=-\log q_i +\overline{\log q_i},
\end{equation*}
in which $\EE=-\overline{\log q_i} = -\sum q_i\log q_i$ is the traditional definition of system entropy. Thus, $\EE_i^*$ is the deviation of the entropy in the $i$th dimension from the system entropy. The constant $\Gx=\EE-\Gl\zbar$ is absorbed by expressing $\EE_i^*$ and $z_i^*$ as deviations from their average values. The constant $\Gk$ is given by \Eq{kappa}, in which $\Gs_{\Gfh}$ is the standard deviation of the forces, $\Gfh_i^*=\EE_i^*-\Gl z_i^*$.

The constraint $\obq\cdot\bz=B$ implies 
\begin{equation*}
  \Gl=\Gb_{{\scriptscriptstyle\EE}z} - \frac{B}{\Gk \Gs^2_z}.
\end{equation*}
The term $\Gb_{{\scriptscriptstyle\EE}z}$ is the regression coefficient of $\EE_i$ on $z_i$, which transforms the scale for the forces of constraint imposed by $\bz$ to be on a common scale with the direct forces of entropy, $-\log\bq$. The term $B/\Gk \Gs^2_z$ describes the required force of constraint on frequency changes so that the new frequencies move $\zbar$ by the amount $\obq\cdot\bz=B$. The term $\Gs^2_z$ is the variance in $\bz$.

When the $\bz$ values change, the changing frame of reference with respect to $\bz$ follows from \Eq{zprice} as $\bq\cdot\obz=\ozbar-B$. When $\zbar$ is a conserved quantity and the $\bz$ values remain constant such that $\bq\cdot\obz=0$, then $\ozbar=B=0$. When $B=0$, the force of constraint for the conserved quantity is expressed simply by $\Gl=\Gb_{{\scriptscriptstyle\EE}z}$. 

\section{Equilibrium thermodynamics and probability}

This section analyzes how the system equilibrium arises from the direct force causing maximum increase in entropy and the constraining forces imposed by $\bz$. That equilibrium can be interpreted as the maximum entropy probability distribution.

The dynamics are expressed in \Eq{qdotphi} as $\oq_i=\Gk q_i\Gfh_i^*$. Equilibrium requires that the forces be constant in each dimension, thus $\Gfh_i^*=0$. We can take that condition as the forces in each dimension given by
\begin{equation*}
  \Gfh_i = \log\frac{\qhati}{q_i}=0,
\end{equation*}
which means that the equilibrium condition can be written as $\log q_i=\log\qhati$. We can express $\qhati$ in terms of the system quantities, $\bz$, that set the forces of constraint. From \Eq{logqhat}, we write the equilibrium condition as $\log q_i=\log k-\Gl z_i$, or
\begin{equation*}
  q_i=ke^{-\Gl z_i}.
\end{equation*}
That probability distribution is the classic Jaynesian thermodynamic equilibrium\autocite{jaynes57information,jaynes57informationII,jaynes03probability} that arises by maximizing entropy subject to a constraint on $\zbar$. That constraint is usually interpreted as a conserved quantity, such that $\ozbar=0$, and $\obq\cdot\bz=\bq\cdot\obz=0$. We can use multiple constraints on a set of system values $\zbar_j$, and replace $\Gl z_i$ by $\sum\Gl_jz_{ij}$ summed over $j$. For simplicity, I focus on a single constraint.

Suppose we want to find a Lagrangian that leads to the Jaynesian equilibrium, in which the defined forces $\bqhat$ arise from a constraint on a conserved system quantity, $\zbar=\bq\cdot\bz=\Gm$. The following Jaynesian Lagrangian does the job
\begin{equation}\label{eq:jaynesL}
  \LL=\EE + \tilde{k}\left(\sum q_i - 1\right) 
    - \Gl\left(\sum q_iz_i - \Gm\right),
\end{equation}
in which $\EE=-\sum q_i\log q_i$, is the classical expression for entropy defined earlier. This Lagrangian is simply the entropy, $\EE$, subject to two constraints. First, the total probability must be one. Second, the system quantity $\zbar=\sum q_iz_i$ is conserved and equal to $\Gm$. The terms $\tilde{k}$ and $\Gl$ are the Lagrangian multipliers that adjust to guarantee that the constraints are satisfied.

Maximum entropy subject to the constraints requires $\prt\LL/\prt q_i=0$, which yields the maximum entropy probability distribution
\begin{equation*}
  q_i=ke^{-\Gl z_i},
\end{equation*}
in which $\log k=\tilde{k}-1$, and $\Gl=1/\Gm$. We can extend this result to unify the commonly observed probability distributions within a single framework by noting that $z_i=\tr(x_i)$ is an arbitrary scaling relation of an underlying value, $x_i$ (\textcite{frank14how-to-read,frank16common}). 

Two conclusions follow. First, equilibrium probability distributions at maximum entropy express the force of constraint on total probability and the forces of constraint on total system quantities. The point of maximum entropy occurs at the minimum relative entropy, $\KL{\bq}{\bqhat}$, which is achieved as $\bq\rightarrow\bqhat$.

Second, pattern follows from the values of $\bz$ that set the forces of constraint and thus the magnitudes of $\bqhat$. How the $\bz$ values arise has not been specified. Thus, the study of pattern often reduces to the study of how various processes set $\bz$. The analysis here clarifies how those processes and the associated maximum entropy probability distribution relate to the universal Price equation expression for the dynamics of populations.

\section{Interpretation of maximum entropy path}

The previous sections analyzed forces in terms of Price's partition of direct and inertial forces, an abstract generalization of d'Alembert's principle of mechanics.  By analogy with d'Alembert's principle, the Price equation term $\obq\cdot\bF$ can be thought of as an abstraction of the virtual work associated with the direct and constraining forces. 

The direct forces are $\bF$. The constraining forces are included in the allowable set of displacements, $\obq$, taken relative to the fixed frame of reference. Such displacements relative to a fixed frame of reference are sometimes called virtual displacements, thus the name virtual work for the term $\obq\cdot\bF$. The Lagrangian expressions provide a method for maximizing the virtual work subject to the constraints that limit the possible set of displacements

We may interpret the partition of direct and constraining forces in different ways, to match the interpretation of different problems. In this article, I split the total direct forces into a direct force that increases entropy, $\bF=-\log\bq$, and a set of potential virtual displacements, $\obq$, that obey the forces of constraint defined by conservation of a functional, $\zbar$, of the system quantities, $\bz$, where one can think of each $z_i$ as a function on the subset, $i$, of the population.

In particular, I defined the total direct forces by $\Gfbh=\log{\bqhat/\bq}$, and then split those forces as
\begin{equation*}
  \obq\cdot\Gfbh = -\obq\cdot\log\bq + \obq\cdot\log\bqhat 
    = -\obq\cdot\log\bq -\Gl\,\obq\cdot\bz.
\end{equation*}

If we take $\Gfbh$ as the direct forces, then the frequency changes can be obtained from the Lagrangian in \Eq{lagr1} that maximizes the action $\obq\cdot\Gfbh$, which is equivalent to minimizing the change in relative entropy, $\KL{\bq}{\bqhat}$. 

If we take $-\log\bq$ as the direct forces, then the frequency changes can be obtained from the Lagrangian in \Eq{lagr2} that maximizes the action $-\obq\cdot\log\bq$, which is equivalent to maximizing the gain in entropy, $\EE$. 

In other words, the realized path maximizes the production of entropy when analyzed within the fixed frame of reference, thus the maximum entropy production principle. That conclusion holds only in the d'Alembert-Price distinction between direct and constraining forces, in which we choose to interpret all direct forces except entropy production as constraining forces on the possible virtual displacements, $\obq$. In addition, the changes in frame of reference that typically arise from change in location, $\obq$, or from change in the constraining forces, are separated by the Price equation approach into the consequences of the inertial forces. 

Maximum entropy production only holds for the partial change from the direct forces, when separating all direct forces other than entropy into the constraints, and when ignoring changes in the frame of reference associated with the inertial forces.

Does it make sense to follow this particular partition of forces into components? There is no correct answer to that question. The principle exists. The interpretations of usefulness and meaning will always have a strongly subjective aspect. 

I follow \textcite{lanczos86the-variational} in the claim that separating direct, inertial, and constraining components is the great unifying perspective in the study of forces. In many systems, it makes sense to describe most of the applied forces in terms of the constraining forces of conserved system quantities. Often, all that remains is the only truly universal force, the increase of entropy, which completes the description of the total direct forces acting on a system. 

In some cases, it may make sense to use a different partition of applied forces into direct and constraining component forces. When the remaining direct component of force differs from entropy alone, then it would appear that the system does not follow the maximum entropy production principle. However, it is better to say that the maximum entropy production principle always holds, but alternative expressions may provide a more meaningful perspective for particular problems.

In this interpretation, entropy is simply a geometric description of position and change for probability distributions when located in logarithmic coordinates. That fundamental geometry explains the universality of entropy, or information, in widely different disciplines and applications.

\section{Geometry and the Fisher information metric}

We can write the conservation of total probability expression in \Eq{fisherdiscrete} for small changes, $\obq$, as
\begin{equation*}
  \obq\cdot\bF + \obq\cdot\bI 
  	= \sum\frac{\oq_i^2}{q_i} - \sum\frac{\oq_i^2}{q_i} = \F_F-\F_I=0,
\end{equation*}
in which $\F=\sum\oq_i^2/q_i$ is the Fisher information metric, and the subscripts denote the direct and inertial components of the Price equation. 

In various models of natural selection, information, and entropy, different measures arise in terms of the Jeffreys divergence, $\J$, the Kullback-Leibler divergence, $\D$, and the Fisher information metric, $\F$. Confusion sometimes occurs, because in the limit of small changes, all three measures converge to an equivalent form that often appears as the Fisher information metric. That limiting equivalence hides the significant differences between the measures and the different situations to which each measure naturally applies. 

The Fisher information metric is used in many applications\autocite{kullback59information,cover91elements}. For example, \textcite{frieden04science} has emphasized that this Fisher information partition subsumes nearly all of the key results of theoretical physics. Similarly, the subject of information geometry subsumes nearly all of the classical aspects of statistical inference through a Riemannian geometry based on the Fisher information metric\autocite{amari00methods}. 

From the general perspective of the Price equation and d'Alembert's form for the conservation of total probability in \Eq{dalemb1}, the partition into Fisher information components arises as a special case in the limit of small changes\autocite{frank15dalemberts}. In that special case of Fisher information, in which $\obq\cdot\bF=\F_F$, one does not separate the forces of constraint from the other directly applied forces. Instead, all directly applied and constraining forces combine into a single quantity that describes the path, in which that path has a natural geometric expression in terms of the Fisher information metric. That geometry is very useful in many applications. But it is important to recognize the more general perspective of Price and d'Alembert, which allows a deeper conceptual understanding of the different roles played by directly applied forces, constraining forces, and inertial forces.

One can think of the maximum entropy production principle in terms of Fisher information geometry. The universal direct force that increases entropy is always present. In addition to that universal direct force, various additional constraining forces combine to influence the curvature of the space of allowable virtual displacements. The direct and constraining forces combine to determine the paths of change within the Fisher information geometry\autocite{amari00methods}.

\section{Direct work, information and entropy}

I summarize in two parts. In this section, I briefly review the Price equation formulation of the work of the direct forces. I then show how the classic measures of information and entropy follow from simple geometric assumptions about the most useful scale on which to measure changes in populations. The following section focuses on the Lagrangian analysis of the dynamical paths of change, including the partial maximum entropy production principle, and provides a final summary.

The Price equation presents universal principles of total change in populations. The strongest principles arise when studying change purely in terms of altered probability distributions. In that case, the natural selection definition of relative fitness as the ratio of probabilities, $1+a_i=w_i=q_i'/q_i$, leads to a Price equation expression for the change in average relative fitness, describing the conservation of total probability in \Eq{consA}, as
\begin{equation*}
    \GD\abar = \bdq\cdot\ba + \bq'\cdot\bda = 0.
\end{equation*}
We can write that conservation law for total probability in terms of d'Alembert's partition of direct, inertial and constraining forces in \Eq{dalemb1} as
\begin{equation*}
    \GD\abar = \left(\bF + \bI\right)\bdq = 0.
\end{equation*}
The allowable displacements in probability, $\bdq$, must obey any constraints imposed on changes in the system, and thus implicitly reflect any underlying forces of constraint. Such displacements may be reversed, because all allowable displacements fall within the constraints of conserved total probability. Reversible infinitesimal displacements that obey the constraining forces, taken in the context of the fixed frame of reference in the initial state of the population, are often called virtual displacements.

In this abstract Price equation generalization of d'Alembert's principle of mechanics for conserved systems, the first component of change arises from the direct forces, $\ba=\bF$, which may be written from \Eq{dqa} as
\begin{equation*}
  \bdq\cdot\bF=\bdq\cdot\ba 
  	= \left(\frac{\bdq}{\bq}\right)\bdq = \sum_i\frac{\left(\GD q_i\right)^2}{q_i},
\end{equation*}
which is the nondimensional product of a displacement multiplied by a force, yielding the Price equation abstraction of the mechanical notion of the work of the direct forces. For infinitesimal displacements, $\bdq\rightarrow\obq$, consistent with the forces of constraint, the term $\obq\cdot\bF$ is often called the virtual work.

The work of the direct forces describes change in the context of the fixed frame of reference given by the initial population. The total change depends on how the frame of reference changes, captured by the second term $\bq'\cdot\bda=\bdq\cdot\bI$, as in \Eq{qpda2}. 

Often, it is difficult to interpret the changing frame of reference in a simple way. Instead, the strongest universal principles come from study of the work of the direct forces---the partial change caused by the direct forces with respect to the fixed initial frame of reference. 

The work of the direct forces may be partitioned into components of directly applied forces, $\bF$, and constraining forces expressed by the allowable displacements, $\bdq$. One can make that partition in a variety of ways according to the interpretation of a particular system. The emphasis on forces helps greatly in understanding the causes of change\autocite{lanczos86the-variational}. 

Fitnesses, $w_i=q_i'/q_i$, are ratios of probabilities. Geometrically, it is convenient to have identical ratios correspond to identical distances between coordinates of probability. We achieve that identity by expressing fitness in logarithmic coordinates
\begin{equation*}
  m_i=\log w_i = \log\frac{q_i'}{q_i}=\log q_i' - \log q_i.
\end{equation*}
When we interpret fitness as a force, the logarithmic coordinates change the multiplication of fitness components of force into the addition of the logarithmic fitness components of force, as in \Eq{wmult}. 

In the Price equation, we can use any arbitrary coordinates, $\bz$, for the quantitative property values associated with probabilities. We can think of those arbitrary coordinates as a geometric transformation of the fundamental coordinates of conserved probability and fitness, $\bw\mapsto\bz$. Equivalently, we may write $\ba\mapsto\bz$, because $\ba=\bw-\bOne$, and the Price equation is shift invariant. 

When we transform from the fundamental coordinates of fitness to the logarithmic coordinates of fitness, $\bw\mapsto\bmm$, we obtain many of the classic expressions for information and entropy, which ultimately express the simple underlying geometry of change described by the Price equation. For example, in logarithmic coordinates, the work of the direct forces becomes
\begin{equation*}
  \bdq\cdot\bF=\bdq\cdot\bmm = \bdq\cdot\GD\log\bq=\J_F,
\end{equation*}
which is the Jeffreys divergence measure of entropy or information, as in \Eq{jeffreys}. The symmetric Jeffreys divergence is the sum of reflected asymmetric Kullback-Leibler divergences, in which the Kullback-Leibler divergence is the most commonly used measure of relative entropy or relative information.

When the changes, $\GD q_i/q_i$, are small, the logarithmic measure of fitness converges to the linear measure of fitness, $\bmm\rightarrow\ba$, and the Jeffreys divergence and the Kullback-Leibler divergence converge to the Fisher information metric. The Fisher metric is the fundamental measure of distance between probability distributions that forms the basis of much of statistical inference and information geometry. 

In these Price equation descriptions of change, we have taken the fitnesses as given, and equated fitness or the logarithm of fitness with a notion of force. That approach is essentially inductive, in which we take the probabilities as given locations, $w_i=q_i'/q_i$, and implicitly induce the force that would be consistent with the change from $q_i$ to $q_i'$. 

\section{Partial maximum entropy production}

The main point of this article is to analyze the traditional deductive perspective of dynamics with respect to force. In that traditional perspective, we begin with the initial location of the population, $\bq$, and given forces which we denote $\bF\equiv\Gfbh$. From those given conditions, we then deduce the changes in location and the new probabilities, $\bq'$. I confined the analysis to the study of small changes, $\obq$. 

To obtain the dynamics, $\obq$, from the initial location and the given forces, I first wrote the Lagrangian expression for each particular case. The Lagrangian focuses on a first term, often called the action, which is either maximized or minimized (extremized). When minimized, the procedure follows the principle of least action, but more generally the procedure is known as the principle of extreme action. 

In this article, I maximized the virtual work of the given direct forces, $\obq\cdot\bF=\obq\cdot\Gfbh$. Intuitively, this simply means that the changes will follow the lines of force in relation to the magnitudes of the force in each dimension. However, we must consider both the direct and constraining force.

The Lagrangian approach provides a natural way to combine direct and constraining forces. In each Lagrangian, the first term gives the virtual work of the direct forces to be maximized. The remaining terms give the constraints that must be satisfied, usually as some total quantity that is conserved when summed over all dimensions of the system. The Lagrangian procedure transforms the system constraints into the constraining force components in each dimension. 

The various results in the text show how different kinds of constraints and different ways of separating overall force into direct and constraining components determine the change in frequencies. 

The key result concerns the partial maximum entropy production principle, which I briefly review. I expressed the given forces as $\Gfbh=\log\bqhat/\bq$. Thus, the virtual work of the given forces in \Eq{split} is
\begin{equation*}
    \obq\cdot\Gfbh=\obq\cdot\log\bqhat-\obq\cdot\log\bq.
\end{equation*}
I assumed that there is some quantity, $\bz$, such as energy or biomass or any other appropriate measure, that is constrained so that the total direct changes in that quantity are $\obq\cdot\bz=B$. We may relabel the part of the given forces, $\log\bqhat$, as a constraining force associated with the fixed value imposed on direct changes in $\bz$, given by the expression in \Eq{zconstrain} as
\begin{equation*}
  \obq\cdot\log\bqhat=-\Gl\lr{\obq\cdot\bz} = B.
\end{equation*}
With this component labeled as a constraining force, the remaining part of the virtual work of the direct forces is $-\obq\cdot\log\bq$, which in the limit for small changes is the production of entropy along the path of small changes, $\obq$. This component is the action term maximized along the path of change, thus the path follows the direction that maximizes the production of entropy. I call this the partial maximum entropy production principle, because the result expresses the change in terms of the fixed frame of reference of the initial population state. Total change must also evaluate any changes in the frame of reference through the inertial forces.

The entropy production principle simply expresses the basic geometry for the path of change when extrinsic forces are considered as constraints on system quantities, and logarithmic coordinates are used to locate populations. Because changes in probabilities as fitness or force have a natural expression as the ratio of probabilities, $w_i=q_i'/q_i$, and such quantities combine multiplicatively, logarithmic coordinates arise naturally from the transformation that yields additive combinations. Thus, entropy production or changes in information arise as the inevitable consequence of the geometry of change when evaluated in the Price equation partition of direct and inertial forces.

In summary, several different disciplines share the same basic fundamental theory of change. From the perspective of the Price equation, we have seen common expressions for natural selection, aspects of physical mechanics and thermodynamics, entropy expressions for probability distributions, and common measures of information theory. Perhaps many common models of learning by reinforcement\autocite{sutton98reinforcement,szepesvari10algorithms} and Bayesian updating\autocite{shalizi09dynamics,harper10the-replicator,campbell16universial} will also share the same underlying geometric principles.

\section*{Acknowledgments}

\noindent National Science Foundation grant DEB--1251035 supports my research.  I did this work while on fellowship at the Wissenschaftskolleg zu Berlin.

\bigskip
\bibliography{main}

\section{Appendix: Literature in specific disciplines}

\subsection*{Natural selection}

Price originally formulated his equation as an expression of natural selection\autocite{price70selection,price72extension}. In another article, without any direct connection to the Price equation, he speculated about a unified theory of change based on an abstract generalization of the principle of selection\autocite{price95the-nature}. 

In Price's vision for a general theory of selection, he suggested the separation of frequency and property values in the description of population change. He also described changes by an abstract mapping scheme between members of two populations. Price never connected these abstract ideas about mapping and about separating frequency and property directly to his formulation of the Price equation, although one can see hints of this in \textcite{price72extension}. 

In other work\autocite{price72fishers}, Price clarified one of the great puzzles in the history of evolutionary theory. In 1930, \textcite{fisher30the-genetical} stated his fundamental theorem of natural selection as: ``The rate of increase in fitness of any organism at any time is equal to its genetic variance in fitness at that time.'' 

Fisher emphasized the exactness of the theorem and his belief that the theorem was a general and profound statement about natural selection. The puzzle is that Fisher's theorem holds exactly only under a very restricted set of assumptions\autocite{crow70an-introduction}. Fisher is regarded as perhaps the greatest mathematical biologist ever. So the mismatch between Fisher's strong claim and the seemingly obvious failure of the theorem was hard to reconcile.

Price\autocite{price72fishers} solved the puzzle. In the language of the present article, Fisher meant that the rate of increase in fitness equals the variance in fitness when evaluated with respect to the fixed frame of reference of the population's initial state. Selection acts as a direct force, with consequences of the direct force evaluated by holding constant the context. Any changes to the population that alter the fitnesses of individuals are regarded as consequences of inertial forces that alter the frame of reference.

Price\autocite{price72fishers} did not use the language of direct and inertial forces, but he clearly understood Fisher's partition of total change into two components. Later work clarified a variety of early theories about natural selection within the context of the Fisher's partition\autocite{ewens89an-interpretation,ewens92an-optimizing,frank92fishers}.

In summary, Price left three separate insights about natural selection: the Price equation, the separation of frequency and property in an abstract mapping scheme, and Fisher's method of partitioning total change with respect to the frame of reference. My own work has unified those different pieces into an extended, more general and abstract interpretation of the Price equation\autocite{frank95george,frank97the-price,frank12naturalb,frank12naturalc}. 

Another important line of work in evolutionary theory concerns the path of change in gene frequencies. \textcite{wright31evolution,wright32the-roles} initiated the approach most closely related to analogies with classical mechanics. That line of work continues to be developed, including explicit connections to notions of entropy and statistical mechanics\autocite{vladar11the-contribution}. 

The studies initiated by Wright contrast with Fisher's approach\autocite{frank12wrights}. In the language of this article, Fisher emphasized instantaneous change at a point and the partition of direct and inertial components of change. Fisher believed that the inertial components of change were too unpredictable to allow an explicit theory for the full path of change over significant lengths. By contrast, Wright and his descendants sought a theory of the paths of change over significant distances. This article emphasized the Fisherian perspective.

\subsection*{Maximum entropy production}

Jaynes' theory of maximum entropy\autocite{jaynes57information,jaynes57informationII,jaynes03probability} emphasizes that probability distributions can be read as expressions of constraining forces \autocite{frank14how-to-read}. 

For example, a Gaussian distribution expresses a constraint on the average distance of observations from the mean value. If one constrains that average distance of fluctuations from the mean, then the Gaussian distribution arises by maximizing the entropy subject to that constraint. Maximizing entropy is roughly equivalent to minimizing information or maximizing randomness. 

Jaynes' maximum entropy describes an equilibrium condition\autocite{jaynes57information,jaynes57informationII,jaynes03probability}. The idea is that entropy increase is a ubiquitous force---a ubiquitous entropic force. Increasing entropy plus constraining forces together define the form of the equilibrium distribution. 

The increase of entropy toward an equilibrium leaves open the problem of the dynamical path followed from initial condition to final equilibrium state. What characterizes the increments along that path? One possibility is that each increment follows the direction that maximizes the increase in entropy---the path of maximum entropy production (MEP). 

Some authors have proposed MEP as a fundamental principle similar to the principle of least action\autocite{dewar05maximum,14beyond}. By that view, essentially all realized paths of motion maximize the production of entropy. Other authors have suggested that MEP is only an approximate description of dynamics\autocite{14beyond}. By that view, certain special systems follow MEP exactly, whereas many other systems follow MEP approximately or not at all. 

The logical status of MEP as a principle and its usefulness in analysis remain open problems. The interpretation of MEP is important, because that interpretation reflects our general understanding of diverse subjects and the relations between those subjects.

In this article, I showed that MEP is an exact statement about dynamics when interpreted in the context of the Price equation and the information theory definition of entropy. The Price equation provides an abstraction of change that may be interpreted as a partition into components that separate direct, inertial and constraining forces.

This Price equation separation of forces is an abstract generalization of d'Alembert's principle of classical mechanics\autocite{lanczos86the-variational}. The Price equation formulation can be applied to both conservative and nonconservative systems, extending d'Alembert's application to conservative systems. Wang\autocite{wang07from} proposed a different way to connect entropy and d'Alembert through a more traditional thermodynamic approach.

Although MEP is a valid principle, I suggested that a purely geometric interpretation provides a more fundamental and universal perspective than does the entropy perspective of MEP. In particular, the conservation of total probability imposes strong geometric symmetry and constraint on the separation of direct and inertial forces\autocite{frank15dalemberts}. Maximum entropy production is a useful but often unnecessarily complicated way of expressing those fundamental geometric principles. 

Returning to Jaynes, his goal was to express an abstract and general approach to understanding probability patterns. He sought to transcend the specific physical assumptions of statistical mechanics and thermodynamics, thereby achieving a more general theory that applied to broader range of disciplines. 

In several ways, Jaynes did not go far enough. For example, he retained entropy and information as primary quantities. Similarly, information geometry, based on metrics such as Fisher information, retains a notion of information as primary. In my view, the underlying geometry, conserved quantities, and symmetries provide the true foundation for analysis as, for example, in \textcite{frank16common}.

\subsection*{Statistical inference and learning algorithms}

This article showed that natural selection connects to universal expressions of population change and probability through the Price equation\autocite{price70selection,price72extension,frank95george,frank12naturalb}. One can think of natural selection as an algorithm for accumulating information. Many authors have noted formal connections between natural selection, information theory\autocite{frank09natural,frank12naturalc}, Bayesian updating in statistical inference\autocite{shalizi09dynamics,harper10the-replicator,campbell16universial}, and learning algorithms\autocite{campbell74evolutionary}. 

Although initial connections have been made between natural selection and those different subjects, unification based on a deeper geometric foundation remains an open problem. For example, Jaynes maximum entropy approach ultimately aimed to unify probability, information, statistical inference, and physical theories of statistical mechanics and thermodynamics\autocite{jaynes03probability}. Another subject which might eventually coalesce is reinforcement learning\autocite{sutton98reinforcement,szepesvari10algorithms} which provides the basis for aspects of neuroscience, cognitive science, and machine learning. 

How do those various subjects relate to general underlying geometric principles for the dynamics of change in populations?

\end{document}